# Energy flows in the earthquake source before and after the main shock

A.V. Guglielmi, O.D. Zotov

*Schmidt Institute of Physics of the Earth, Russian Academy of Sciences; Bol'shaya Gruzinskaya str., 10, bld. 1, Moscow, 123242 Russia; guglielmi@mail.ru (A.G.), ozotov@inbox.ru (O.Z.)*

**Abstract:** We proceeded from general physical concepts based, on the one hand, on the Umoff–Poynting theorem, and on the other, on the phenomenological theory of earthquakes, and formulated the following question: What are the directions of energy flows in the earthquake source before and after the formation of a main rupture in it? A non-standard technique for experimental research of this issue has been developed. The epicentral zone of the main shock is considered as a kind of track detector, and foreshocks and aftershocks are considered as marks (tracers) marking the propagation in the source of some factor that has energy and stimulates the excitation of foreshocks and aftershocks in a stressed-strained rock mass. By processing and analyzing a large volume of observation data, it was found that over time, foreshocks, on average, approach the epicenter of the main shock, while aftershocks, on the contrary, move away from the epicenter. A method is indicated for verifying the result by studying the magnitude dependence of foreshock convergence and aftershock divergence.

## Introduction

Over the past decade, a small team of geophysicists has developed the fundamentals of the systematics and phenomenology of tectonic earthquakes [1–14]. The systematics, which, as is known, consists of classification and nomenclature, is based on the idea of classical, mirror and symmetrical triads formed by foreshocks, main shock and aftershocks. The phenomenology is based on a differential evolution equation containing quadratic nonlinearity. Both methodological approaches turned out to be very effective. In this work, we continue this line of development of earthquake physics and study the question of the directions of energy flow in the source before and after the formation of a main rupture in the continuity of rocks. Note that the concept of energy transfer in continuous media was introduced by Umoff in 1874 in his doctoral dissertation "The Equation of Energy Motion in Bodies". In earthquake physics, Umoff's idea, as far as we know, is used for the first time.

Let us explain the essence of phenomenology using the example of the simplest nonlinear equation of evolution

$$\frac{dn}{dt} + \sigma n^2 = 0. \tag{1}$$

Here $n$ is the frequency of aftershocks, $\sigma$ is the source deactivation coefficient [4]. The equation was proposed based on guiding considerations, and its applicability was tested experimentally. The equation has two important properties. Firstly, it allows us to pose the inverse source problem, i.e. calculate the deactivation coefficient based on aftershock frequency data. By solving the inverse problem, it is shown that Omori's law [15] is satisfied only at the first stage of the evolution of aftershocks, called the Epoch of Harmonic Evolution (EHE). It was found that at the end of the EHE, a bifurcation occurs and the state of the earthquake source changes radically.

Secondly, the form of equation (1) suggests interesting generalizations. A nontrivial generalization is the logistic equation

$$\frac{dn}{dt} = n(\gamma - \sigma n), \tag{2}$$

where $\gamma$ is the second phenomenological parameter of the theory [9]. Equation (2) allowed us to see a rich collection of phase portraits of the earthquake source.

By adding the diffusion term to the logistic equation

$$\frac{\partial n}{\partial t} = n(\gamma - \sigma n) + D\nabla^2 n, \tag{3}$$

we obtain an equation that describes the spatiotemporal evolution of tremors. In mathematics, the nonlinear diffusion equation is known as the Kolmogorov–Petrovsky–Piskunov equation [16]. It has the remarkable property that its solutions contain traveling nonlinear waves. Waves of aftershock activity propagating from the epicenter of the main shock were first observed in [6].

It is quite clear that aftershocks themselves do not move, but serve as a kind of markers (tracers) that mark the propagation of some factor in the source that has energy and stimulates the excitation of aftershocks in the stressed-deformed body of the source. This consideration led us to the question: If after the main shock the flow of energy is directed away from the epicenter, then wouldn't we expect that before the main shock the flow is directed towards the epicenter? In other words, our heuristic hypothesis is that before the main shock, the projection of the Umoff vector onto the earth's surface is directed towards the epicenter of the main shock. The hypothesis seems quite plausible, since the main shock occurs as a result of energy accumulation in the vicinity of the hypocenter. Nevertheless, the hypothesis needs experimental verification.

## Observations

We want to highlight a subtle effect and for this we will need a special technique for processing and analyzing observational data. After a series of tests, one of the authors (O.D.) developed a non-standard technique, the essence of which is as follows.

The general idea is that the epicentral zone of the main shock is considered as a kind of track detector, and foreshocks and aftershocks are considered as tracers, marking the direction of energy flow. To detect a weak effect against a background of numerous and practically uncontrollable noise, we need information on a large number of earthquakes. We used data contained in the USGS/NEIC earthquake catalog from 1973 to 2019. 138 events were identified. The event was considered to be an interval of $\pm 100$ days relative to the day of the main shock with a magnitude of $M \geq 7.5$ and depth of the hypocenters of foreshocks, main shock and aftershocks of no more than 250 km. All foreshocks and aftershocks in the epicentral zone of the main shock were taken into account. The radius of the epicentral zone for each event was calculated using the method described in [10]. A total of 1960 foreshocks and 34627 aftershocks were observed.

Data processing consisted of a sequential series of procedures. First of all, the events were synchronized at the time of the main impact. Then the foreshocks and aftershocks were ordered in a somewhat unusual way. Namely, they were all combined into clusters, each of which was characterized by the serial number of the appearance of a foreshock or aftershock in each individual event. It is assumed that clustering smoothes out sharp differences in the geological conditions of occurrence of specific events. Finally, for each cluster, the average distance between the epicenters of foreshocks or aftershocks and the epicenter of the main shock was calculated using formula $r = \sqrt{x^2 + y^2}$. Here $x$ and $y$ are Cartesian coordinates on the earth's surface, and the center of the coordinate system is combined with the epicenter of the main shock.

The result of processing foreshocks and aftershocks is presented in the figure in the left and right parts, respectively. The cluster numbers are shown along the horizontal axes. Distances $r$ are plotted along the vertical axes. We see that over time, aftershock activity on average moves away from the epicenter of the main shock.

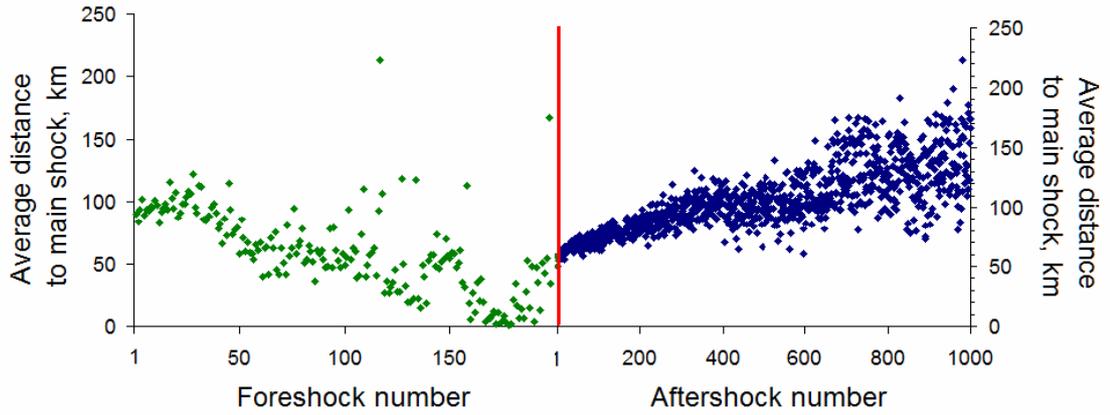

The effect of foreshock convergence and aftershock divergence (see text).

This independently confirms the trend first discovered in [6]. In contrast, foreshock activity approaches the epicenter of the main shock. The result corresponds to the expectation that was expressed in the Introduction when setting the problem.

**Discussion**

The result described above is qualitative, since the method of data processing and analysis does not allow us to calculate the values of the migration velocity of foreshocks and aftershocks. However, on a qualitative level, we can state with a high degree of confidence that the activity of foreshocks, on average, approaches the epicenter of the main shock, while the activity of aftershocks moves away from the epicenter. If the points in the figure are approximated by straight lines, then for foreshocks and aftershocks we will have

$$r[\text{km}] = -0.39\,j + 97.8 \qquad (4)$$

and

$$r[\text{km}] = 0.08\,j + 67.9 \qquad (5)$$

respectively. Here $j$ is the cluster number. The corresponding correlation coefficients are $-0.67$ and $0.78$.

Judging by the figure, the converging flow of foreshocks is directed towards the epicenter of the main shock. This is consistent with the idea of energy flow from the surrounding space to the hypocenter, where the main rupture of rocks begins, leading to the main shock. As for aftershocks, judging by the figure, the diverging flow begins not at the epicenter, but at some distance $\Delta r$ from it. Perhaps the reason for the displacement of 70 km is the unloading of stress in the vicinity of the hypocenter after the formation of a main rupture. Our hypothesis can be tested experimentally as follows.

It is known [17] that the length of the fault $L$, on which the radius of the unloading zone depends, is greater, the higher the magnitude of the main shock: $\log L[\text{cm}] = 3.2 + 0.5M$. For example, $L = 90$ km at $M = 7.5$, and $L = 160$ km at $M = 8$. Let us pay attention to the fact that the indicated values of $L$ are comparable to the value of $\Delta r \approx 70$ km. Based on our interpretation of displacement $\Delta r$, then when selecting events according to criterion $M \geq 6.5$, the displacement will be approximately half as large as what we see in the figure.

**Conclusion**

We considered the epicentral zone of the main shock as a kind of track detector, and foreshocks and aftershocks as markers, presumably marking the movement of Umoff's energy flows in the source. It was previously found that aftershock activity, on average, moves away from the epicenter. From general considerations, this observation led us to the hypothesis that foreshock activity, on the contrary, approaches the epicenter. To test the hypothesis, an original technique for processing observation data has been developed. Analysis of the processing results confirmed the existence of foreshock convergence and aftershock divergence. We pointed out the possibility of critical testing of our ideas by analyzing the magnitude dependence of the displacement of the point from which aftershock migration begins relative to the epicenter of the main shock.

We express our sincere gratitude to B.I. Klain and A.D. Zavyalov for numerous discussions and constant assistance in the work. We thank colleagues at the US Geological Survey for lending us their earthquake catalogs USGS/NEIC for use. The work was carried out according to the plan of state assignments of Schmidt Institute of Physics of the Earth, Russian Academy of Sciences.

This study will be presented in an expanded form to the journal "Geodynamics and Tectonophysics".